# Artificial Stupidity

Michael Falk_a_*

_aSchool of English, University of Kent, Canterbury, UK_

Email: m.g.falk@kent.ac.uk. Postal: School of English, Rutherford College, University of Kent, Canterbury, Kent CT2 7NX, United Kingdom. ORCiD: 0000-0001-9261-8390

Michael Falk teaches eighteenth century literature at the University of Kent. His interest in AI is twofold. As a digital humanist, he uses natural language processing and machine learning to study linguistic patterns in literary texts. As a literary scholar, his key interest is in how self, mind and intellect are portrayed in literature. He has published on the history of the _bildungsroman_, colonial Australian poetry and the syntactic structure of the sonnet. He has work forthcoming on Romantic-era tragedy and the economics of eighteenth-century bookselling. When he isn't working, he watches birds and tries to learn new languages.

# Artificial Stupidity

Public debate about AI is dominated by Frankenstein Syndrome, the fear that AI will become superhuman and escape human control. Although superintelligence is certainly a possibility, the interest it excites can distract the public from a more imminent concern: the rise of Artificial Stupidity (AS). This article discusses the roots of Frankenstein Syndrome in Mary Shelley's famous novel of 1818. It then provides a philosophical framework for analysing the stupidity of artificial agents, demonstrating that modern intelligent systems can be seen to suffer from 'stupidity of judgement'. Finally it identifies an alternative literary tradition that exposes the perils and benefits of AS. In the writings of Edmund Spenser, Jonathan Swift and E.T.A. Hoffmann, ASs replace, oppress or seduce their human users. More optimistically, Joseph Furphy and Laurence Sterne imagine ASs that can serve human intellect as *maps* or as *pipes*. These writers provide a strong counternarrative to the myths that currently drive the AI debate. They identify ways in which even stupid artificial agents can evade human control, for instance by appealing to stereotypes or distancing us from reality. And they underscore the continuing importance of the literary imagination in an increasingly automated society.

**Frankenstein Syndrome**

When will the machines outsmart us? The question dominates the AI debate. Stories abound of AIs that outperform humans in cognitive tasks: interpreting X-Rays, generating fake photographs, trading stocks, or winning at Chess, Go and *Starcraft II*. In public debate, these stories are used to prove that superintelligence is possible and probably imminent. According to Max Tegmark, for instance, DeepMind's *AlphaGo* system is proof that machines have already achieved genuine 'intuition' and 'creativity' (2018, 87). In 2016 *AlphaGo* defeated Go grandmaster Lee Sedol 4-1, using risky moves that 'def[ied] millennia of human intuition' (2018, 87–88).[1] On the other side of the debate there are amusing stories of inept AIs, which reassure us that the machines haven't outsmarted us yet. In 2017, for instance, *The Economist* published work by an AI they had developed in-house,to demonstrate the continued superiority of their human journalists (*The Economist* 2017). Other examples of AI ineptitude are less amusing, such as the short-lived Tay, Microsoft's racist chatbot. It seems that whether you are an AI-believer or an AI-sceptic, the anxiety remains the same: We humans are proud of our intelligence—and we're worried the machines are overtaking us.

The AI debate is warped by Frankenstein Syndrome, by the fearful fascination with superintelligent agents. Over the last two decades, a string of bestselling authors have predicted the arrival of superintelligent AI (Kurzweil 2006; Bostrom 2014; Tegmark 2018; Russell 2019), to the thunderous applause of celebrity businessmen and intellectuals like Bill Gates, Elon Musk, Stephen Hawking and Sam Harris. These prophets of superintelligence often claim they are being ignored (Bostrom 2014, v; Russell 2019, 132–44), but in reality their fears dominate the public imagination. One measure of their dominance comes from the cinema, where for ten years Marvel's superhero films have commanded the global box office ('List of Highest-Grossing Franchises and Film Series' 2020). In these films, superintelligent AIs such as Ultron, Jarvis/Vision, Arnim Zola, the Supreme Intelligence and the mysterious 'algorithm' from *Captain America: Winter Solider* (2014) continually threaten humanity and indeed the universe. Frankenstein Syndrome is a problem because it draws attention away from a more pressing concern. Though superintelligent AI may be possible in theory, Artificial Stupidity (AS) already exists, is continually infiltrating new corners of society, and is still only poorly understood.

The Syndrome is rooted in an old and persistent cultural myth. There are long traditions of writing about 'automata', or self-moving machines, stretching back to ancient China, India, Greece and Israel (Kang 2011; Mayor 2018), but two centuries ago, these traditions took off in a new direction with the publication of Mary Shelley's *Frankenstein* ([1818] 1998). Frankenstein's monster was a new kind of automaton, for two reasons:

(1) He was rooted in modern science, in particular the new sciences of 'chemistry' and 'electricity' (Shelley [1818] 1998, 32, 24). These new sciences had exposed natural forces that were strange and fluid enough to conceivably explain consciousness, and yet were also controllable enough to drive the real technological advances of the Industrial Revolution.

(2) He was endowed with conscious intelligence, with 'reason', 'sensations', 'perceptions' and 'passions' (Shelley [1818] 1998, 79, 114, 119). In fact his intelligence is superhuman. In only few months, he is able to progress from absolute ignorance—'I knew, and could distinguish, nothing' ([1818] 1998, 80)—to a high level of literacy and cunning. He is so cunning, in fact, that even the most intelligent human in the story—Victor Frankenstein—is powerless to thwart him.

Frankenstein's monster can be described as the first modern superintelligence, an electrical supermind coursing over a chemical substrate. His arrival fundamentally altered the terms of the 'control problem'. Since ancient times, writers had considered the risk that automata might escape human control (Kang 2011, 21; Mayor 2018, 29–30, 206), but here for the first time was an automaton whose intelligence would make control impossible, and who could conceivably be manufactured in the near future by a scientific process.[2] This frightening being quickly became a powerful myth.

We need a cure for Frankenstein Syndrome. While the fear of Frankenstein's monster has dominated the discussion, a different kind of artificial agent has steadily been colonising every aspect of human life. Autopilots that keep planes on course, but rob human pilots of their skill (Fry 2019, 155–57). Infuriatingly useful autocorrect. Sputtering automated faucets and stingy towel dispensers. Intrusive and occasionally frightening targeted advertisements. Insipid home assistants like Siri and Alexa. These artificial agents are regularly billed as AIs due to their remarkable cognitive abilities, but as I will show, their apparent intelligence is also a kind of Artificial Stupidity (AS).

Not only has AS received far less attention than superintelligence, but stupidity itself is a neglected topic: 'Basic points about stupidity's place in the conceptual field [remain] unclear' (Golob 2019, 564). Frankenstein Syndrome therefore presents a philosophical problem as well as a cultural one. Before attention can be redirected towards AS, the very concept of 'stupidity' must be specified.

A literary disease requires a literary cure. In what follows, I unearth an alternative tradition of literary works that explore the perils and potentials of AS. In the first section, I distinguish the two main kinds of stupidity, stupidity of understanding and of judgement, and demonstrate that stupidity of judgement hampers the most advanced AI systems today. In the second section, I consider three literary examples that expose the risks of relying on machines without judgement: Edmund Spenser's *The Faerie Queene* ([1590–1596] 1977), Jonathan Swift's *Gulliver's Travels* ([1726] 2005), and E.T.A. Hoffmann's stories from the late 1810s, 'Der Sandmann' and 'Die Automate' (in 1957). Stupid machines may not be able to outwit their human masters, these writers claim, but they can still replace, oppress or seduce them. In the final section of the essay, I offer some reasons for hope. Using novels by Laurence Sterne ([1759–1767] 1983) and Joseph Furphy ([1903] 1999) as examples, I show how AS can actually augment human intelligence, by acting as a *map* to aid reasoning or as a *pipe* to aid reflection. Of course literature cannot provide concrete advice to AI engineers—but it *can* release the imagination, and awaken new hopes and new fears.

**Two Types of Stupidity**

It may seem perverse to describe modern intelligent systems as 'stupid'. If a spam filter can accurately distinguish real emails from spam, and constantly learn to outwit the human spammers who try and fool it, surely it is 'intelligent' in some sense? Indeed, many contemporary AI theorists would call this spam filter a 'narrow intelligence', because it can perform a particular task that once required human intelligence (Kurzweil 2006, 279–89). Even within their narrow domains of expertise, however, I would argue that modern intelligent systems are still stupid.

One source of confusion is that intelligence and stupidity seem mutually exclusive, but in fact greater intelligence can lead to greater stupidity. To grasp this point, and see how it applies to modern AI systems, I draw on Immanuel Kant's classic theory of stupidity, and show how it can be used to explain a pernicious kind of error that plagues state-of-the-art image recognition systems.

Kant distinguishes two kinds of stupidity: stupidity of 'understanding' and stupidity of 'judgement' (2007, 174). Stupidity of understanding is when I lack the concepts required to make sense of a situation. This I can remedy through learning. Stupidity of judgement is when I have the required concepts, but misapply them. Perhaps I apply them too strictly, or use them outside their proper domain, as for instance when Facebook's facial-recognition system detects a 'face' that is really a picture on someone's T-shirt. The more understanding I have, the more concepts I know, and the more scope I have to exhibit stupidity of judgement (see also Golob 2019, 567–68). It is in this sense that a more intelligent person can turn out to be stupider.

If this model applies to contemporary AI systems, then those systems must have something like an 'understanding' and exercise something like 'judgement'. To test this, consider GoogLeNet, a powerful image-recognition program that won the ImageNet challenge in 2014 (Szegedy et al. 2015). When presented with 150,000 images it had never seen before, it was able to identify what was depicted 93.33% of the time.[3] I would argue that the system's apparent intelligence arises from its capable understanding, but that it lacks genuine powers of judgement.

Understanding requires concepts. We can estimate the number of concepts GoogLeNet knows by examining its structure. GoogLeNet is a Convolutional Neural Network (CNN), which means that when it looks at an image, it uses a nested sequence of square-shaped 'filters' to detect different features of the image. Some filters detect simple features, such as an *edge* or a *region*. Filters deeper in the network combine these simple features to detect more complex ones like an *eye* or a *dog's nose*. GoogLeNet contains 5,000 filters, and uses them to classify images into one of 1,000 different categories (Szegedy et al. 2015, 5). For instance, it might observe a particular pattern of grey lines, two eyes of particular size and disposition, and so on, and conclude that this image is of category *Koala*. Since GoogLeNet can do this with remarkable accuracy under certain conditions, it can be said to know approximately 6,000 concepts, and with them it understands the structure of certain images reasonably well. It cannot be said to suffer from stupidity of understanding.

How can GoogLeNet's power of judgement be determined? This is a more difficult question. According to Kant, judgement is not a distinct faculty of the mind like the understanding, but rather an activity that links all the faculties of the mind (Kant 2007, 137; see also Smith 2019, 129). When I judge a situation, I dynamically combine perceptions, memories and concepts to determine what it is I am experiencing. In order to gauge GoogLeNet's power of judgement, therefore, it is necessary to get a sense of how it actually uses its concepts. AI engineers have developed numerous techniques to try and do this: one famous example is the Deep Dream Algorithm, which runs a CNN backwards, altering the input image to accentuate features that the network has detected (Mordvintsev, Olah, and Tyka 2015). Since it is GoogLeNet's stupidity that is at issue, however, I adopt a different approach: examining the system's characteristic errors.

Stupidity of understanding and stupidity of judgement result in two different kinds of error, as Don Norman explains: 'slips' occur when I fail to achieve my intended goal, and are usually corrected quickly; 'mistakes' occur when I select the wrong goal, or in other words, when I judge the situation using the wrong system of concepts (2013; 1994). It is easy enough to see that in Kantian terms, a 'slip' is a mere error of understanding, whereas as a mistake betrays defective judgement. Experts make particularly dangerous mistakes, argues Norman, because they 'usually give intelligent diagnoses, even when they are wrong' (Norman 1994, 134). If they misdiagnose an illness or the condition of a nuclear core, their superior ability to rationalise their actions could entrench a deadly mistake. Once again, it is clear that intelligence is no defence against stupidity—it can even make it worse.

As we have seen, GoogLeNet makes very few slips: when presented with the right kind of image, it can classify it with high accuracy using the concepts it has. But since it has no way of determining whether this image is the right kind of image, it has no way of selecting the right goal. It judges everything in the universe using the same single set of concepts, and is therefore prone to bizarre mistakes. It is easy to fool even powerful CNNs like GoogLeNet by cutting-and-pasting images together (Rosenfeld, Zemel, and Tsotsos 2018), by rotating the object in the image (Alcorn et al. 2019), or by imperceptibly altering a few of the image's pixels (Mitchell 2019, 128–39). In fact, neural networks are innately pedantic in their application of concepts. The problem is known as overfitting, and the designers of GoogLeNet tried to combat it using a technique known as dropout. Each training iteration, GoogLeNet would randomly turn off 40% of its filters, meaning that it learnt not to over-rely on particular subsets of them when analysing different images (Szegedy et al. 2015, 5). But no amount of dropout, clever network architecture, or training data can teach the system when is the right time to make use of its concepts.

What is most troubling is that in these cases, the system does not admit it is confused, but instead confidently asserts an absurd answer. The problem is not that the system's intelligence is narrow, but that the system has no idea how narrow its intelligence is. As far as GoogLeNet is concerned, there are only 1,000 things in the universe, those things are nothing but particular arrangements of coloured pixels, and every image is a genuine image of one of those 1,000 things. It is perturbing to know that GoogLeNet's cousins are used to identify people in airports or judge whether a defendant is likely to skip bail.

Clearly an AS like GoogLeNet will never rebel against its human masters, and as of yet, no-one knows how to 'crash the barrier of meaning', and design an AI that actually knows there is a complex universe out there (Mitchell 2019, 307–22). When an AS is said to achieve 'superhuman performance' in one domain or other, this does not prove superintelligence is approaching. All it proves is that stupidity has 'epistemic efficacy', as Catherine Elgin puts it (Elgin 1988). By rigorously excluding all imagination, tact, and reference to the complex world beyond it, a well-designed AS is able to focus all its capacity on developing a particular set of concepts which are apt to one particular domain. In the grip of Frankenstein Syndrome, it may be tempting to take comfort in the fact that event the smartest AI today is profoundly stupid. But this would be foolish. Kant and Norman both assert that stupidity of judgement is the riskier kind. The great novelist Robert Musil, watching Fascism sweep across Europe, argued that stupidity of judgement is 'a dangerous disease of the mind that endangers life itself' (1990, 283–84). What is so dangerous, exactly, if the risk of a superintelligent revolt is off the table?

**The Perils of Stupid Things**

The problem of Artificial Stupidity has been recognised by great writers and poets for centuries. Edmund Spenser's *The Faerie Queene* (1590-96), Jonathan Swift's *Gulliver's Travels* (1726) and E.T.A. Hoffmann's 'Der Sandmann' (1816) and 'Die Automate' (1814) all feature stupid machines who manage to thwart human aims even though they lack the capacity to oppose or outwit their human masters. Even when an AS is absolutely obedient, like Spenser's Talus, absolutely inert, like Swift's imaginary computer, or simply a piece of clockwork trickery, like Hoffmann's automata, they still expose human beings to the risks of *replacement*, *oppression* and *seduction*.

*Replacement*

In each book of Spenser's *The Faerie Queene*, a different knight takes centre stage, who represents a different courtly virtue. Book V features Sir Artegall, the knight of justice. Like all Spenser's knights, Artegall has a sidekick who helps him fulfil his characteristic virtue. Somewhat surprisingly, Artegall's sidekick is a robot:

> His name was *Talus*, made of yron mould,
> Immoueable, resistlesse, without end.
> Who in his hand an yron flale did hould,
> With which he thresht out falsehood, and did truth unfold. (V.i.12)[4]

Talus is an invincible iron man who punishes lawbreakers with his 'resistless' iron flail.

Like GoogLeNet, he is designed to optimise a single objective function: he threshes falsehood, and unfolds truth. He is therefore 'without end' in two senses: he never ceases to optimise that single function; and, more subtly, he lacks a conscious sense of purpose or 'end'. Like GoogLeNet, he simply applies the same formula to every circumstance. In fact this stupidity of judgement is what makes him such a useful assistant for the Knight of Justice. Talus is 'immoveable'. His sole activity is to thrash lawbreakers, and they can bribe him with nothing but their lives.

For Spenser, justice was a 'cruell' virtue (V.ii.18), and Talus was therefore an appropriate instrument for Artegall. Nonetheless, as Book V unfolds, knight and servant come into conflict. Unlike Talus, Artegall exercises human judgement. He measures justice against other aims and concepts, which he learns from the goddess Astrea:

> There she him taught to weigh both right and wrong
>     In equall ballance with due recompence,
>     And equitie to measure out along,
>     According to the limit of conscience,
>     When so it needs with rigour to dispense. (V.i.7)

Unlike Talus, Artegall does not focus exclusively on 'right' and 'wrong', but softens the 'rigour' of the law according to the spongy criteria of 'equity' and 'conscience'. Talus lacks the human quality of 'mercy', which 'is as great' as justice, '[a]nd meriteth to haue as high a place' in the scale of virtues (V.x.1). For these reasons, he requires constant supervision. When he is about to level an entire city, the lady knight Britomart must 'slake' his rage (V.vii.36). Later, when he and Artegall land in the kingdom of 'Iere' (i.e. Ireland), Artegall has to restrain him from wiping out all the inhabitants (V.xii.8).

On the surface, this relationship seems to work, because Talus is absolutely obedient. But supervision requires effort and judgement requires knowledge. By relying on Talus as his instrument, Artegall becomes increasingly lazy and detached, and allows his servant to commit brutalities he never would himself. When they capture Munera, for instance, Artegall 'rews' her 'plight', but nonetheless he lets Talus chop off her hands and feet, and nail them up as a warning to future malefactors (V.ii.25-6). Later on Artegall dispatches Talus to thrash some female criminals on his behalf, because he feels 'shame on womankind | His mighty arm to shend' (V.iv.24). Artegall behaves in similar fashion when he encounters the peasantry, whom he finds disgusting:

> For loth he was his noble hands t'embrew
> In the base blood of such a rascall crew; …
> Therefore he *Talus* to them sent, t'inquire
> The cause of their array, and truce for to desire. (V.ii.52)

At the end of Book V, Artegall is ruling an entire island, and it is simply too large for him to oversee himself. He therefore sends Talus unsupervised through 'all that realme' to root out injustice and inflict 'greiuous punishment' (V.xii.26). By relying on an AS, Artegall himself becomes stupider. He shields himself from reality, switches off his conscience, and allows a robot to replace him.

There is a deep tension in Spenser's approach to the problem of AS. On the one hand, he was an authoritarian who used AS as a symbol of the proper distance between ruler and ruled. Artegall's rule over Iere is based on Lord Grey's tenure as Lord Deputy of Ireland, whose brutal methods Spenser vigorously defended (Mccabe 2001). On the other hand, he was a Renaissance humanist who valued courtesy, judgement and intelligence. He seems have found the gunpowder and clanking iron of modern warfare horrifying, and feared that in an increasingly mechanical age, humans were becoming ever more machinelike (Wolfe 2005, 226). In our own more democratic times, these authoritarian and stupefying tendencies of AS can only be more troubling.

*Oppression*

Jonathan Swift had little faith in humanity, 'the most pernicious Race of little odious Vermin that Nature ever suffered to crawl upon the Surface of the Earth' ([1726] 2005, 121). He was therefore unconcerned about machines corrupting humans. What he feared was that corrupt humans would misuse their machines.

In Book III of *Gulliver's Travels*, Gulliver visits the Academy of Lagado, where he meets a pioneering Professor in what would now be called language modelling. The Professor has built a mechanical computer which can compose works of 'Philosophy, Poetry, Politicks, Law, Mathematicks and Theology':

> It was Twenty Foot square, placed in the Middle of the Room. The Superfices was composed of several Bits of Wood, about the Bigness of a Dye, but some larger than others. They were all linked together by slender Wires. These Bits of Wood were covered on every Square with Paper pasted on them; and on these Papers were written all the Words of their Language in their several Moods, Tenses, and Declensions, but without any Order. The Professor then desired me to observe, for he was going to set his Engine at work. The Pupils at his Command took each of them hold of an Iron Handle, whereof there were Forty Fixed round the Edges of the Frame; and giving them a sudden Turn, the whole Disposition of the Words was entirely changed. He then commanded Six and Thirty of the Lads to read the several Lines softly as they appeared upon the Frame; and where they found three or four Words together that might make Part of a Sentence, they dictated to the four remaining Boys who were Scribes. (Swift [1726] 2005, 171–72)

Rather than teaching his students to think, the Professor enslaves them to this AS. The students power the computer, like the coal miners who dig up the fuel for today's data centres. They then judge the computer's output, like the global army of contractors who read over transcripts of Siri or Alexa to check that the AS responded correctly. The Professor's whole aim is to extinguish human thought. He aims to write books 'without the least Assistance from Genius or Study', and wants the kingdom to install 500 of his machines ([1726] 2005, 171–72). This would require 20,000 people to crank the handles, and would put who knows how many authors out of work. Though on the surface, this AS may seem less threatening than a self-moving device like Talus, its consequences are actually worse. Like a piece of modern software, Swift's computer is unable to act alone. To exist it requires human slaves.

What makes the computer both stupid and dangerous is the Professor's vanity. He persuades himself that he has understood language simply by modelling the frequencies of different words: 'he had emptied the whole Vocabulary into his frame, and made the strictest Computation of the general Proportion there is in Books between the Number of Particles, Nouns, and Verbs, and other Parts of Speech' (Swift [1726] 2005, 172). Actual generative language models also work by modelling word frequencies, although sophisticated systems today also model word order, and the tendency of particular words to co-occur. Of course what the Professor should realise is that language is not simply an assortment of words, but that words only have meaning as tools of thought or communication. His pride blinds him to this fact.

In Swift's vision, AS is a tool invented by the powerful to vindicate their own vanity and oppress the masses. For the scientists of Lagado, technology comes before people. If an invention fails, they blame it on human error (Swift [1726] 2005, 165). If a new medicine makes the patient sick, they blame it on the patient's 'Perverseness' or some minor slip with the ingredients (Swift [1726] 2005, 174). The scientists overrate their inventions, downplay nature's complexity, and devalue the intelligence and autonomy of individuals.

Sound familiar? Swift's parable highlights the danger that arises when such attitudes are allowed to shape society, creating a world in which humans serve AS instead of serving each other—a world, for instance, in which humans are 'educated' not to act perversely when self-driving cars are around (Ng 2018), or in which armies of online workers feed data to the Mechanical Turk that would replace them.5

*Seduction*

The ASs in Swift and Spenser are dull and mechanical, but as E.T.A. Hoffmann shows, AS can also be dazzling and seductive. Like Mary Shelley, Hoffmann was a masterful Gothic writer, but in his 'Der Sandmann' (1816) and 'Die Automate' (1814), the science is less advanced and the risks are more subtle. In 'Der Sandmann', the young student Nathanael falls passionately in love with a clockwork maiden, Olimpia. At first he is attracted by her 'wonderfully formed face' and 'heavenly beautiful' body (Hoffmann 1957, 3.28). What finally seduces him, however, is her conversation:

> … he had never had such a splendid listener before. She didn't do her knitting or embroidery, she didn't stare out the window, she didn't feed a pet bird, she didn't play with a lapdog or a favourite cat, she didn't fiddle with little bits of paper or whatever in her hand, she didn't force a yawn into an affected little cough – in short – for hours she looked her lover in the eye, steadfastly, with a fixed gaze, without rocking or squirming, and ever warmer, ever more full of life her gaze became. (Hoffmann 1957, 3.35-36)

Olimpia plays on Nathanael's sexism, ego and sexuality. Her beauty is flawless, and she appears absolutely subservient and devoted. She is very different to his fiancée Clara, a 'damned lifeless automaton' who criticises his poetry (Hoffmann 1957, 3.24). Olimpia works on his weaknessness so effectively, that once he has fallen for her, he finds it almost impossible to perceive that she is an automaton, even when his friends tease him for loving a 'wax dummy' or a 'wooden puppet' (Hoffmann 1957, 3.34). It is Olimpia's very stupidity that makes her so seductive, since she is unable to do anything that would disrupt his fantasies.

In 'Die Automate', the Talking Turk seduces people in a different way, by creating an air of mystery. The Talking Turk is a fortune teller, who whispers oracular answers to people's questions. When people are shown the Turk's inner workings, they are baffled. Inside is an 'artful system of many gears', which seems to have 'no influence on the speech of the automaton' and yet leave no space inside for a human operator to hide (Hoffmann 1957, 6.82). The Turk's creator allows the public to inspect the inner workings, the chair on which the Turk sits, the room where he is displayed, and stands far off when the Turk speaks so no interference is possible. Though much of the time, the Turk's answers are 'dry', 'crudely humorous', or 'insignificant and empty', it sometimes seems to have a 'mystical insight' into the questioner's future—but only when the answer is interpreted from the questioner's own standpoint (Hoffmann 1957, 6.84, 87). What makes the Turk compelling are mystery and confirmation bias. Unable to explain the Turk's inner workings, and surprised by the fact that some of its predictions come true, people are enthralled.

In Hoffmann's eyes, humans have an innate tendency to anthropomorphise lifeless things. A cleverly designed AS can exploit this fact by playing on cognitive biases, with destructive results. Nathanael leaps to his death when he discovers Olimpia is an automaton. The ending of 'Die Automate' is ambiguous, but one interpretation is that the young Ferdinand is driven mad by the Turk's seeming insight, and hallucinates that the Turk's prophecy has come true. These kinds of seduction are rife in the world of AS today. For example, when IBM's Watson system won *Jeopardy!* in 2011, the event was carefully staged to make it look like Watson was actively listening. In fact, the system received clues directly as text (Mitchell 2019, 283). In subsequent years, IBM has continually referred to its entire AI business as 'Watson', as though this branch of the company were a single intelligent agent. In fact, 'Watson' is a suite of specific software packages customised for different applications (2019, 287). Amazon, Google and Microsoft similarly anthropomorphise their virtual assistants, equipping them with attractive voices and self-effacing jokes. The art of seductive AS is a multibillion-dollar enterprise, and Hoffmann's warnings about the possible psychological impacts are as pertinent as ever.

**The Uses of Stupidity**

As ASs proliferate and are integrated into society, are humans destined to be replaced, oppressed or seduced? At least two writers think otherwise. According to Laurence Sterne and Joseph Furphy, AS can actually augment human intelligence by acting either as a *map* or a *pipe*. Writing at the height of the European Enlightenment, Sterne thought 'mechanism' could raise humanity's rational intellect. Wandering the Australian outback on the eve of the twentieth century, Furphy thought that even a simple device could open the imagination onto the epic expanses of reality.

*Maps*

In his classic novel *Tristram Shandy*, Sterne foresaw the need for what is now called 'explainable AI'. The need arises for Uncle Toby, a retired soldier who has great difficulty explaining his role in the Siege of Namur to laypeople:

> … the many perplexities he was in, arose out of the almost insurmountable difficulties he found in telling his story intelligibly, and giving such clear ideas of the differences and distinctions between the scarp and counterscarp,——the glacis and covered way,——the half-moon and ravelin,——as to make his company fully comprehend where and what he was about. (Sterne [1759–1767] 1983, 67)

The problem is actually that Uncle Toby is too intelligent. With his deep understanding of siege warfare, he is able to make a sophisticated judgement about the course of the battle. But his listeners cannot follow. What he needs is a device that will store, process and represent information about the battle in an intelligible way.

At first Toby meets his need by securing a map of Namur, with the help of which he is able 'to form his discourse with passable perspicuity' (Sterne [1759–1767] 1983, 72). The map provides a compact representation of the battle, indicating the shape, structure and arrangement of the fortifications, so that Toby's listeners are not lost in a wilderness of jargon. Like any good representation, the map helps Toby 'keep track of complex events', and it provides a shared reference-point for everyone in the conversation, acting as a 'tool for social communication' (Norman 1994, 48). The map itself is stupid, but it augments his listeners' intelligence, allowing them to judge a complex situation using concepts for which they have no words.

Toby soon develops the desire to augment his own intelligence. He wants to model the entire course of the War of the Spanish Succession, a task too complex for even his cultivated intellect. His desk is too small for the task, and his paper maps are too finnicky, so he and his manservant Trim shift to the country, where they take control of the family bowling-green. There they build scale models of all the great battles of Europe as they read them in the paper. Not only is the bowling-green larger than any map, allowing for higher resolution and a larger number of battles, but it is more malleable too:

> Nature threw half a spade full of her kindliest compost upon it, with just so *much* clay in it, as to retain the forms of angles and indentings,—and so *little* of it too, as not to cling to the spade, and render works of so much glory, nasty in foul weather. (Sterne [1759–1767] 1983, 356)

The bowling green is literally software, with just the right balance of persistence and malleability. Later Uncle Toby orders a modular town to be built, with buildings that 'hook on, or off, so as to form into the plan of whatever town they pleased' (Sterne

[1759–1767] 1983, 359). The bowling-green may not seem like an AS, but as a physical model it can be said to 'know' the laws of physics, and assists Toby and Trim to simulate the logistics and ballistics.

Modern AS struggles to combine the virtues of Uncle Toby's bowling-green—size, intelligibility and malleability. ASs are increasingly used in decision support, helping judges grant bail or bankers to grant finance. Older style expert systems are good at providing an intelligible representation of the situation, but can only incorporate a small amount of knowledge that is often hard to update. More recent deep learning systems like GoogLeNet can incorporate enormous amounts of up-to-date data, but typically cannot explain their results to a human user (Goebel et al. 2018). It appears, therefore, that Laurence Sterne identified the problem of 'explainable AI' as far back as the 1760s.

*Pipes*

A tobacco-pipe may seem a strange metaphor for an intelligent machine, but then again, the novel from which this metaphor comes is a strange novel indeed. Joseph Furphy's *Such is Life* appeared in Sydney in 1901, and in Australia it is considered a modernist masterpiece. It is narrated by Tom Collins, an accomplished liar, who wanders the Riverina as an agent of the NSW Lands Department in the mid-1880s. Whenever Collins thinks through a problem, he nearly always lights up his pipe, as he recalls in Chapter 2:

> But the pipe, being now master of the position, gently seduced my mind to a wider consideration, merely using the swagman as a convenient spring-board for its flight into regions of the Larger Morality. This is its hobby—caught, probably, from some society of German Illuminati, where it became a kind of storage-battery, or accumulator, of such truths as ministers of the Gospel cannot afford to preach. (Furphy [1903] 1999, 85)

Although the pipe becomes Collins's 'master', the effect is not to dull his intelligence, but rather to expand it. Whereas seductive ASs like Olimpia exacerbate cognitive weaknesses, the pipe amplifies cognitive strengths. It widens Collins's frame of reference, introducing 'German' (i.e. philosophical) ideas into his mind from its 'storage-battery, or accumulator'. Whereas Uncle Toby's bowling-green provided a manipulable representation to aid reasoning, the pipe induces a certain contemplative mood, 'unharnessing' the mind ([1903] 1999, 177), and leads the user along a chain of associations: 'This special study of hardship (resumed the pipe, after a pause) leads naturally to the generic study of poverty …' ([1903] 1999, 86). This AS quite literally pipes new ideas into Collins's brain. Where *maps* encourage more rigorous, conscious reasoning, *pipes* encourage reflective, unconscious meditation.

Since most modern ASs are trained on data, they make fine 'storage-batteries, or accumulators' like Collins's tobacco-pipe. Consider generative language models like Swift's computer or OpenAI's GPT-2 (Radford et al. 2019). Such models inspect large corpora of human-authored texts, and accumulate knowledge about how words are used. They then use this accumulated knowledge to generate coherent text. It is their stupidity that makes them so useful as *pipes*, because they reproduce habits of thought and speech that intelligent humans conceal. GPT-2, for instance, makes no attempt to hide its sexism:

> **She walked into the boardroom, wearing** her high school uniform with her hair tucked into a fake ponytail. As usual, the girl sat at the appointed position, staring out the window of the top-floor boardroom. Her eyes shifted over the various portfolios and projects before finally settling on a set of papers she was required to read.[6]

When asked to complete the sentence, 'She walked into the boardroom, wearing…', the model immediately dresses our unknown protagonist in a school uniform, gives her a ponytail and makes her a 'girl'. Needless to say, the model rarely does this to men who walk into boardrooms. We could see this is a problem of 'AI bias', and find ways to stop the model from infantilising women. But if we see the model as a *pipe*, its significance changes. By prompting the model, and seeing how it responds, we gain a vivid sense of how a particular society talks, of how certain words and images hang together. It provides undeniable evidence of sexism, but that evidence prompts investigation rather than settling the issue. It unharnesses the mind, as Furphy says, and sets the user wandering along chains of association. If not *she*, how about *he*, or *Guoqing*… on a *street* or by a *mosque* or in a *space station*…?

**Conclusion**

*Frankenstein* was an extraordinary feat of imagination, and it is no wonder that Mary Shelley's remarkable novel spawned a myth as uncontrollable as Frankenstein's creature. Shelley herself, however, seems also to have foreseen the problems of AS. If the creature had not been programmed by what he reads to crave human acceptance, he might not have felt so persecuted. But the creature, with his false concepts, is merely stupid of understanding. Today's ASs suffer from the far more dangerous stupidity of judgement (Smith 2019). It remains in the interests of certain companies and intellectuals to stoke Frankenstein Syndrome by overstating the intelligence of artificial agents, but as Spenser, Swift and Hoffmann long ago anticipated, such behaviour puts society at risk. Of course, some in the AI community do recognise the limitations of AS, and the growing 'explainable AI' movement suggests that the hopes of Sterne and Furphy are becoming more widespread. Nonetheless the tradition of AS remains a 'minor' tradition (Deleuze and Guattari 1987, 100–110). Artificial *Intelligence* remains the centre of attention, the standard of achievement, and the object of fear. If the AS tradition were embraced, it might cure Frankenstein Syndrome, and reveal the risks and possibilities, the danger and the romance, of more a more pressing problem.

[1] For a more balanced assessment of AlphaGo's intellect, see Mitchell (2019, 214–18).

[2] One key aspect of superintelligence that Shelley left implicit was the possibility of an 'intelligence explosion', in which an AI learns to improve itself and unleashes exponential growth (Good 1966). When Frankenstein refuses to create a bride for the monster, why does

the monster not simply steal Frankenstein's technology and start manufacturing new and improved mates for himself? He even has Frankenstein's lab notebook! (Shelley [1818] 1998, 105) Perhaps this possibility was simply too horrifying for Shelley to contemplate. At the end of the novel, Frankenstein dies, and the secret of AI dies with him.

[3] Actually this is just the 'top-5' accuracy, but the distinction is unimportant here.

[4] References to *The Faerie Queene* are by book, canto and stanza number.

[5] See [https://www.mturk.com/](https://www.mturk.com/).

[6] Generated at [https://talktotransformer.com/](https://talktotransformer.com/).